\documentclass[twocolumn,prl,sort&compress,floatfix]{revtex4-1}
\usepackage{amssymb}
\usepackage{amsmath}
\usepackage{mathrsfs}
\usepackage{bm}
\usepackage[dvipsnames]{xcolor}
\usepackage[sort&compress]{natbib}

\usepackage{graphicx}

\usepackage{color}
\definecolor{red}{rgb}{1,0,0}
\definecolor{blue}{rgb}{0,0,1}

\newcommand{\red}[1]{\textcolor{black}{#1}}

\begin{document}
\title{Epigenetic Transitions and Knotted Solitons in Stretched Chromatin}

\author{D. Michieletto$^{1,*}$, E. Orlandini$^2$, D. Marenduzzo$^1$}
\affiliation{$^1$ SUPA, School of Physics and Astronomy, University of 
	Edinburgh, Peter Guthrie Tait Road, Edinburgh, EH9 3FD, UK. $^2$ Dipartimento di Fisica e Astronomia and Sezione INFN, Universit\'a di Padova, Via Marzolo 8, Padova 35131, Italy\\ $^*$corresponding author: davide.michieletto@ed.ac.uk }

\begin{abstract}
	\textbf{The spreading and regulation of epigenetic marks on chromosomes is crucial to establish and maintain cellular identity. Nonetheless, the dynamic mechanism leading to the establishment and maintenance of tissue-specific, epigenetic pattern is still poorly understood. In this work we propose, and investigate  {\it in silico}, a possible experimental strategy to illuminate the interplay between 3D chromatin structure and epigenetic dynamics. We consider a set-up where a reconstituted chromatin fibre is stretched at its two ends (e.g., by laser tweezers), while epigenetic enzymes (writers) and chromatin-binding proteins (readers) are flooded into the system. We show that, by tuning the stretching force and the binding affinity of the readers for chromatin, the fibre undergoes a sharp transition between a stretched, epigenetically disordered, state and a crumpled, epigenetically coherent, one. We further investigate the case in which a knot is tied along the chromatin fibre, and find that the knotted segment enhances local epigenetic order, giving rise to ``epigenetic solitons'' which travel and diffuse along chromatin. Our results point to an intriguing coupling between 3D chromatin topology and epigenetic dynamics, which may be investigated via single molecule experiments.}
\end{abstract}

\maketitle

\section{Introduction }

Cell-line-specific features in multi-cellular organisms are achieved by regulation of ``epigenetic marks'', biochemical modifications of DNA and histone octamers which do not affect the underlying genomic sequence. Thus, the average pattern of epigenetic marks in a given-cell line 
well correlates with the pattern of genes which are transcriptionally active or inactive within that cell-line~\cite{Alberts2014,Waddington1968,Waddington1942,Turner2002,Henikoff2016,Zhang2015a}. Dissecting the biophysical mechanisms leading to the {\it de novo} establishment, spreading and maintainance of epigenetic marks is consequently a key step towards a better understanding of the dynamic organisation of genomes and of chromosomal re-arrangement throughout the cell cycle~\cite{Naumova2013}, in cellular ageing~\cite{Pal2016} and pluripotency~\cite{Yamanaka2010}. Notwithstanding their pivotal role, these mechanisms are still poorly understood\red{~\cite{Cortini2016}}.


The epigenetic patterning of chromatin -- the fibre made of DNA wrapped around histone proteins~\cite{Alberts2014} -- has been shown to strongly correlate with the three-dimensional (3D) nuclear organisation of interphase chromosomes~\cite{Rao2014,Beagrie2017,Barbieri2012,Brackley2016nucleus,Jost2014,Dixon2012,Brackley2016nar}. For instance, transcriptionally active regions can be co-localised in multi-enhancer hubs~\cite{Beagrie2017} or transcription factories~\cite{Cook2001book}, whereas transcriptionally inactive regions may form large heterochromatic {\it foci}~\cite{Gilbert2004}, mega-base (Mb) size lamin-associated domains~\cite{Guelen2008} or Barr~\cite{Pinter2012} and Polycomb bodies~\cite{Wani2016}. \red{This intimate connection is also further supported by mean replication timing data~\cite{Yaffe2010,Baker2012,Ryba2010,Boulos2015,Boulos2014, Julienne2013,Boulos2013}.}

On the other hand, it is important to realise that the establishment of epigenetic patterns is fundamentally a {\it dynamic} process, where biochemical tags are constantly deposited, removed and degraded on histones, which can themselves be displaced during transcription or replaced after replication~\cite{Ramachandran2015,Zentner2013,Probst2009,Saksouk2015,Zhang2015a}.   
For this reason, simple models where epigenetic marks are stably deposited along chromatin~\cite{Sexton2012,Filion2010,Rao2014,Jost2014,Barbieri2012,Brackley2016nucleus} are only crude approximations of a much more complex and dynamic scenario. 
Crucially, these ``static'' models fail to address key questions such as, how epigenetic patterns are first established along chromosomes, and how these change, for instance, with cellular ageing~\cite{Pal2016} or during disruptive events in the cell cycle~\cite{Zentner2013}. In addition, an understanding of cell-to-cell variability in genome organisation~\cite{Nagano2013,Nagano2016} and of efficiency of the induced-pluripotency pathway~\cite{Yamanaka2007} may be achieved only through models which can take into account the plasticity of the epigenetic landscape.

In recent years, wide-spread technological advances in the field of molecular biology allowed the biophysical community to identify some of the key players in the dynamics of epigenetic patterning~\cite{Henikoff2016,Talbert2006,Zentner2013,Pal2016}. At the heart of this process are proteins which can ``read'' and ``write'' biochemical tags along chromatin~\cite{Zentner2013,Zhang2015a,Willyard2017}. Importantly, some of these proteins are found in the same complex~\cite{Zentner2013} or are known to directly recruit one another~\cite{Muir2016}. For instance, the heterochromatin binding protein HP1 (a reader) possesses a chromodomain recognising tri-methylation of Lysine 9 on histone 3 (H3K9me3)~\cite{Lachner2001,Hiragami2016,Kilic2015}; at the same time, it can recruit the methyltransferase complex SUV39h1~\cite{Zentner2013} (a writer), therefore engaging a positive feedback loop which self-perpetuates this repressive mark~\cite{Muir2016,Narlikar1,Narlikar2}.

Computational models of this positive feedback loop in 1D~\cite{Dodd2007,Dodd2011} have shown that accounting for long-range contacts is necessary to allow spreading of repressive marks. On the other hand, 3D models coupling epigenetic and chromatin dynamics~\cite{Michieletto2016prx} have shown that spreading of silencing marks on a mobile 3D fibre can be viewed as an ``all-or-none'' transition, where a local fluctuation triggers an epigenetic wave of repressive marks which stabilises a compact globular state~\cite{Michieletto2016prx}. 

Existing experimental and computational studies suggest that the feedback between 3D chromatin structure and epigenetic dynamics along the chromosome may be a key potential mechanism underlying the establishment of epigenetic patterns which can display memory of their landscape. However, direct experimental observations of dynamical epigenetic and chromatin conformational changes in the nucleus are largely missing, in view of the difficulty to achieve enough spatio-temporal resolution {\it in vivo}. In light of this, here we propose, and investigate {\it in silico}, a novel avenue to directly observe the coupling between epigenetic dynamics and 3D chromatin folding. 

The system we consider can be recreated {\it in vitro} via single-molecule experiments: it is inspired both by experiments studying the stretching of reconstituted chromatin~\cite{Kruithof2009,Meng2015,Cui2000}, and by investigations of protein-DNA and protein-chromatin interactions {\it in vitro}~\cite{Kanke2016,Stigler2016}.
Specifically, we envisage a set-up where reconstituted chromatin is attached at both ends to macroscopic beads 
so as to exert stretching forces on the strand via optical or magnetic tweezers~\cite{Kruithof2009,Cui2000}. Further, we imagine that the chromatin fibre is embedded in a solution where reader and writer proteins are added and activated. In practice, one may achieve this by including ``readers'' such as heterochromatin HP1 proteins~\cite{Canzio2013} (which are known to be multivalent, hence act as bridges which fold chromatin), and ``writers'' such as SUV39h~\cite{Zentner2013}. These proteins can respectively read and write the repressive H3K9me3 mark~\cite{Canzio2013,Muir2016}. An alternative option is to use Polycomb PRC complexes and Enhancer-of-zeste (E(z)) proteins which are respectively able to bind and deposit the H3K27me3 mark on histones~\cite{Talbert2006,Angel2011,Zentner2013,Laprell2017,Ciabrelli2017}. 

Our results show that by tuning the strength of the external stretching force $f$ and the binding affinity of reader proteins $\epsilon$, the system can display either a stretched, epigenetically disordered, state (SD) or a compact, epigenetically ordered, one (CO). The two regimes are separated by an abrupt transition line which we quantitatively locate in the $(f,\epsilon)$ parameter space.
We further show that when knots are tied along the chromatin, these can readily be identified by looking at the local order of epigenetic patterning. We argue that the topology of the knots localise and enhance 3D chromatin interactions, thereby ``protecting'' the epigenetically ordered region, which remains localised within the knotted arc. We dub these novel and remarkable states ``epigenetic knotted solitons''.

We finally envisage that our findings will inform the design of novel single-molecule experiments {\it in vitro} and illuminate the interplay between epigenetic dynamics and chromatin topology {\it in vivo}.

\begin{figure*}[t]
	\centering
	\includegraphics[width=0.8\textwidth]{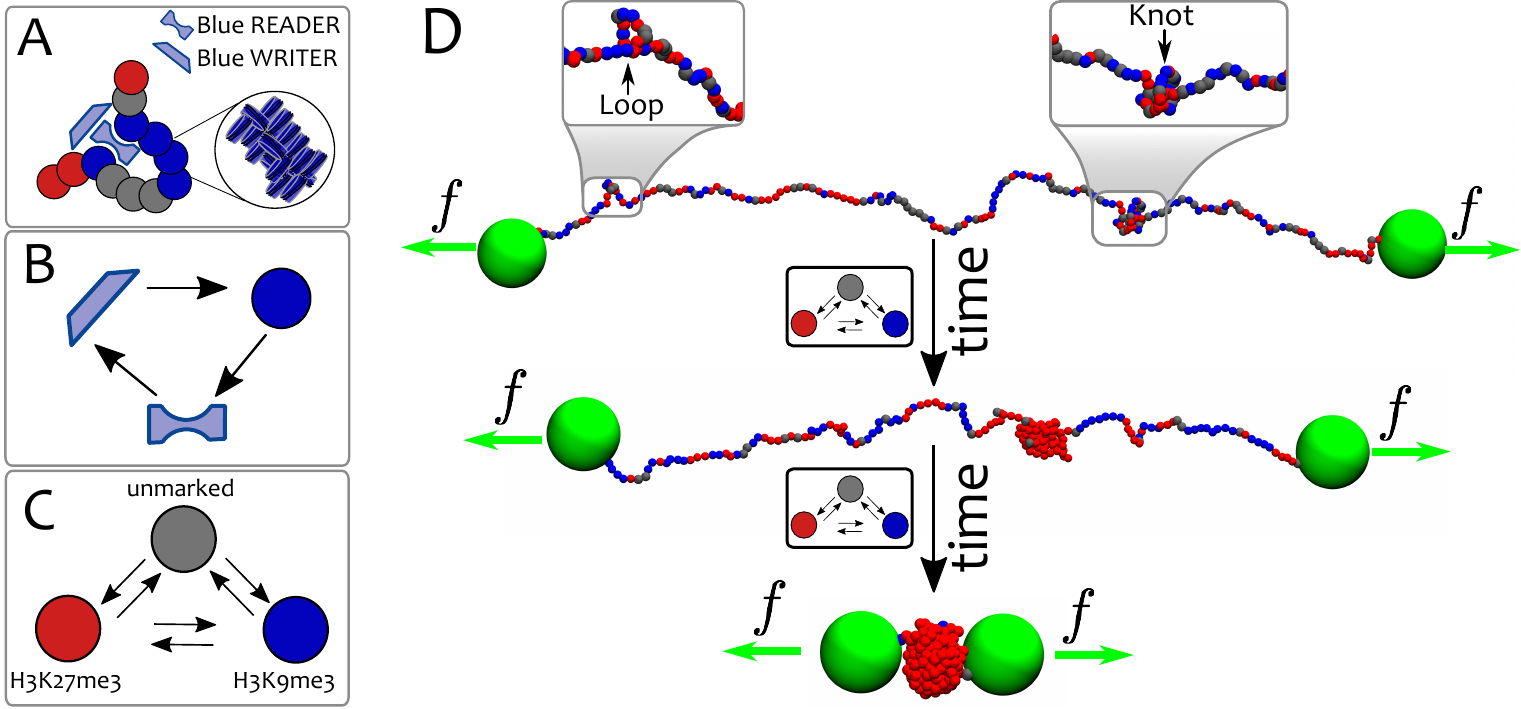}
	\caption{\textbf{Model and set-up.} (\textbf{A}) Chromatin is modelled as a coarse-grained ``recolourable'' bead-spring polymer; each bead represents few nucleosomes and it is coloured either red, blue or grey (unmarked bead). We implicitly model reader and writer proteins interacting with chromatin: the former allow polymer folding by bridging between segments bearing the same colour, the latter deposit new marks on nucleosomes and allow beads to change colour. (\textbf{B}) As red/blue marks are deposited on beads, respective reader proteins can bind to them; in turn, they recruit writer proteins, which deposit new red/blue marks. This mutual recruitment triggers a positive feedback loop. (\textbf{C}) Reader and writer proteins for red and blue marks compete over the same chromatin segments thereby generating dynamic epigenetic marking. (\textbf{D}) Experimental set-up: macroscopic beads (green) are attached at both chromatin ends and allow the strand to be stretched by external tweezers with force $f$~\cite{Kruithof2009,Cui2000}. Starting from a configuration in equilibrium with the external force and with no proteins in solution, the system is flooded with reader/writer proteins thereby triggering the competition between red and blue marks (in the model, this is effectively done by allowing attractive interactions between chromatin segments). The system may eventually evolve towards an equilibrium state where chromatin is compact and only one epigenetic mark dominates ($\epsilon > \epsilon_c(f)$, shown in the figure), or towards one where chromatin is stretched and no coherent mark is established ($\epsilon < \epsilon_c(f)$, not shown).}
	\label{fig:panel_model}
\end{figure*}

\subsection{Chromatin Model and Experimental Set-up}
We model the chromatin as a ``recolourable'' bead-spring polymer (for further details see Methods). Each bead has a size $\sigma$ -- for definiteness, we take $\sigma = 3$ kb or $30$ nm~\cite{Thoma1979,Widom1985,Robinson2006}, but our results do not depend on this precise choice -- and it bears a ``colour'' representing a specific biochemical tag (see Fig.~\ref{fig:panel_model}). For instance, one may think of blue beads as representing chromatin regions with an excess of heterochromatin mark, H3K9me3, and red beads as ones with excess of Polycomb mark, H3K27me3. We further include grey beads which represent unmarked regions of chromatin. \red{For simplicity, in this work we consider only these three colours, in agreement with the generic  experimental observation that few epigenetic ``states'' are sufficient to well capture the overall epigenetic landscape in several organisms~\cite{Roudier2011,Liu2011,Sexton2012,Julienne2013,Julienne2013a}.~\footnote{Extending our model to account for more colours (epigenetic states) do not change the qualitative behaviour of the system.}}  

Beads interact through a Weeks-Chandler-Andersen potential, which strongly suppresses any overlap between beads. This interaction is further modified to account for attractive interactions (with affinity $\epsilon$) between beads bearing the same colour. This effectively mimics the action of reading proteins, or ``bridges'', which can bring together distant chromatin segments bearing the same epigenetic mark. Unmarked (grey) or differently marked beads are therefore considered to be solely sterically interacting. 
The \red{truncated Lennard-Jones (LJ)} potential can be written as a function of \red{distance $x$ between,} and the colour $q$ of, the interacting beads $a$ and $b$ as follows
\begin{equation}
	U_{LJ}^{ab}(x) = \dfrac{4\epsilon_{ab}}{\mathcal{N}} \left[ \left(\dfrac{\sigma}{x}\right)^{12} - \left(\dfrac{\sigma}{x}\right)^6 - \left(\dfrac{\sigma}{x_c^{q_a q_b}}\right)^{12} +
	\left(\dfrac{\sigma}{x_c^{q_a q_b}}\right)^6 \right].
	\label{eq:pot}
\end{equation}
for $x\leq x_c^{q_a q_b}$, whereas $U^{ab}_{LJ}(x)=0$ otherwise \red{($\mathcal{N}$ is a normalisation constant, see Materials and Methods)}. \red{In Eq.~\eqref{eq:pot} $x_c^{q_a q_b}$ is the cut-off distance between beads $a$ and $b$ with colors $q_a$ and $q_b$ respectively.} \red{This potential is a computationally efficient way to model repulsive and attractive interactions between beads. Specifically, it can display an attractive (negative) region by setting the cut-off larger than the minimum of the potential or can yield pure repulsion otherwise. In practice, we use the knowledge of the colours $q_a$ and $q_b$ to define a colour-dependent cut-off as 
\begin{equation*}
x_c^{q_a q_b}=
\begin{cases}
1.8 \sigma \text{ if } q_a=q_b \neq 0 \\ 
2^{1/6}\sigma \text{ if } q_a \neq q_b \text{ or } q_a=0 \text{ or } q_b=0.
\end{cases}
\end{equation*}
}
The shift (last two terms in eq.~\eqref{eq:pot}) ensure that there is no discontinuity in the potential. The binding affinity $\epsilon_{ab}=\epsilon$ when $q_a=q_b \neq 0$ and is a free parameter of our model, otherwise $\epsilon_{ab}=k_BT_L$ when $q_a\neq q_b$ or one of the two is unmarked.   

The ``recolouring'' process is modelled via a Monte Carlo procedure which occurs at an inverse rate of $k_R^{-1} = \tau_{R} = 10^3 \tau_{Br} \simeq 10$ s (see Materials and Methods).
Every time a bead ($a$) is selected for a recolouring attempt, its colour is randomly changed into one of the two remaining colours and the new energy is computed. Because the only colour-dependent potential employed in the simulations is the one written in Eq.~\eqref{eq:pot}, the difference in energy is given by 
\begin{equation}
	\Delta U=\sum_{b}  U_{LJ}^{ab}(x) \, .
\end{equation}  
The recolouring attempt is then accepted with a Metropolis probability 
\begin{equation}
p=\exp{\left(\dfrac{-\Delta U}{k_B T_R}\right)} \, ,
\end{equation}  
where $T_R$ is a generic temperature that can, in principle, be tuned according to the efficiency of the writing process and is therefore independent on the solution temperature $T_L$, which regulates the stochastic dynamics of the beads in 3D~\cite{Michieletto2016prx} (see Materials and Methods).  
\red{The recolouring process and the motion of the chromatin in 3D are engaged in a positive feedback loop: beads that are near each other in space are more likely to become equally coloured and beads bearing the same colour are more likely to stick together. Thus, both species (red and blue) compete over the chromatin strand in a similar way that ``up'' and ``down'' spins compete over a mobile string, or a ``magnetic polymer''~\cite{Garel1999a}.}

\begin{figure*}[t]
	\centering
	\includegraphics[width=0.9\textwidth]{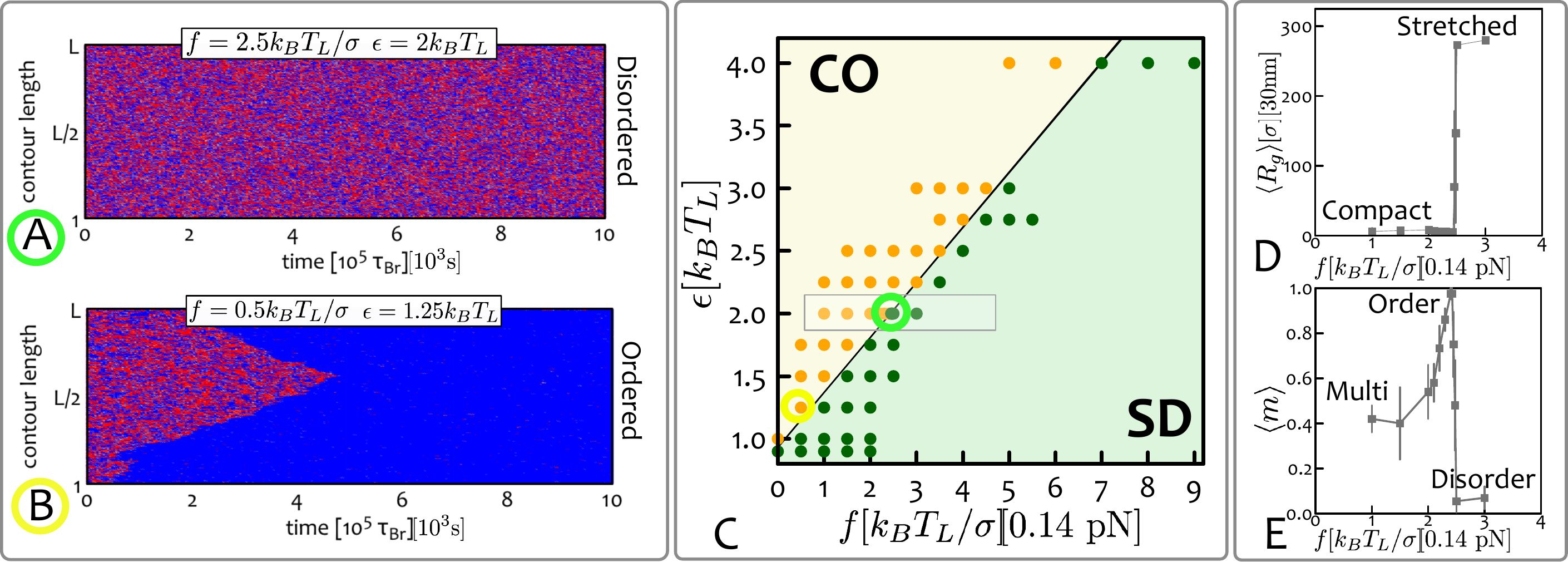}
	\caption{\textbf{Epigenetically-Driven Phase Transition in Stretched Chromatin}. At time $t=0$ the chromatin fibre $L=1000$ beads long is in equilibrium, under a given stretching force $f$. The system is then instantaneously flooded with (implicitly modelled) reader and writer proteins, and the attractive interaction between same-coloured beads turned on. (\textbf{A}-\textbf{B}) The evolution of the system as a function of time can be visualised through ``kymographs'' which show the colour (q) of each chromatin segment (y-axis) as a function of time (x-axis). Depending on the choice of parameters $\epsilon$ and $f$, the system may display (\textbf{A}) stretched-``epigenetically disordered'' (SD) states  or (\textbf{B}) compact-``epigenetically ordered'' (CO) ones.  From the kymographs, one can also appreciate the nucleation and spreading dynamics of epigenetic marks. (\textbf{C}) The phase diagram of the system in the parameter space $(f,\epsilon)$ displays two regions with compact-ordered (CO, yellow shaded) and stretched-disordered (SD, green shaded) equilibrium states. The regions are separated by an abrupt transition line $f_c(\epsilon)$. Data-points corresponding to performed simulations are also shown. (\textbf{D}-\textbf{E}) Show the force dependent of the mean radius of gyration $\langle R_g \rangle$ and the mean absolute magnetisation $\langle m \rangle$ for a fixed value of $\epsilon$ \red{(error bars represent standard deviations)}. Both profiles display an abrupt transition when the critical line $f_c(\epsilon)$ is crossed. Intriguingly, the absolute magnetisation decreases as $f\rightarrow 0$, which can be understood in terms of multiple nucleation points that trigger meta-stable multi-domain states (see below and Materials and Methods). \red{Because these are meta-stable states, the corresponding averages are performed out-of-equilibrium over the last $2$ $10^5 \tau_{Br}$ timesteps}. Data-points for $\epsilon=2 k_BT_L$ and near the transition line have been obtained by averaging over 64 independent replicas. The grey rectangle in \textbf{(C)} highlights the region considered for the profiles reported in panels \textbf{(D-E)}. Mapping to real units of measure are also shown. See also supplementary Movies.
	}
	\label{fig:panel_kymo_pt}
\end{figure*}

Here, we study the behaviour of the system upon tuning the stretching force $f$ and the attraction strength $\epsilon$, since this protocol may be realised {\it in vitro} by using reconstituted chromatin~\cite{Kruithof2009,Cui2000} and proteins such as HP1 and SUV39~\cite{Narlikar1,Narlikar2}, as previously mentioned. For simplicity, we limit to the case $T_L=T_R$ which ensures that the epigenetic read-write mechanism and the chromatin folding are governed by transition rules between different microstates that obey detailed balance and that can be described in terms of an effective free energy. 
Considering out-of-equilibrium conditions~\cite{Michieletto2016prx} leads to transitions between states with similar epigenetic patterns, i.e. from swollen ordered (or disordered) to compact ordered (or disordered), which can be understood as the homopolymer (or heteropolymer) limit of our system. Because these transitions do not shed light onto the interplay between epigenetics and chromatin conformations, we here decide not to pursue them. 
  

In order to quantify the behaviour of the system we perform Brownian Dynamics simulations for a typical runtime of $5$ $10^6$ $\tau_{Br}$ -- corresponding to $1000$ $\tau_R$, or Monte Carlo sweeps -- of a chain $L=1000 \simeq 3$ Mb beads long (if not specified otherwise). Although chromatin has not yet been reconstituted to such a length, the trends we uncover are generic and also hold for smaller values of $L$. 
   
\section{Results}

\subsection{Epigenetic and Conformational Transitions of Stretched Chromatin}

We initialise the systems by assuming that no reader or writer protein is present at $t<0$ and by letting the chromatin equilibrate while subject to the stretching force $f$ and in a bath at temperature $T_L$. In practice, we do this by lowering the cut-off for the Weeks-Chandler-Andersen potential written in Eq.~\eqref{eq:pot} to $2^{1/6}\sigma$ for any pair of beads so that the interactions between beads are purely repulsive. 

At time $t=0$ the system is flooded with reader and writer proteins which can bridge beads with the same epigenetic mark (i.e., colour) and to attempt recolouring of the beads (i.e., writing) at a rate $\tau^{-1}_R$. To visualise the epigenetic evolution of the system, in Fig.~\ref{fig:panel_kymo_pt}(A-B) we report typical ``kymographs'', which show the colour ($q$) of each bead along the polymer at a given time $t$. Kymographs readily capture the epigenetic ``ordering'' of the chromatin strand as a function of time for a specific choice of $\epsilon$ and $f$. 
These two latter quantities are the main free parameters of the system: the stretching force $f$ can be controlled {\it in vitro} through either optical or magnetic tweezers, while the binding affinity between readers and epigenetic marks may be varied by considering mutants of bridging proteins such as HP1~\cite{Narlikar1,Canzio2013}.

 \begin{figure*}[t]
 	\centering
 	\includegraphics[width=0.85\textwidth]{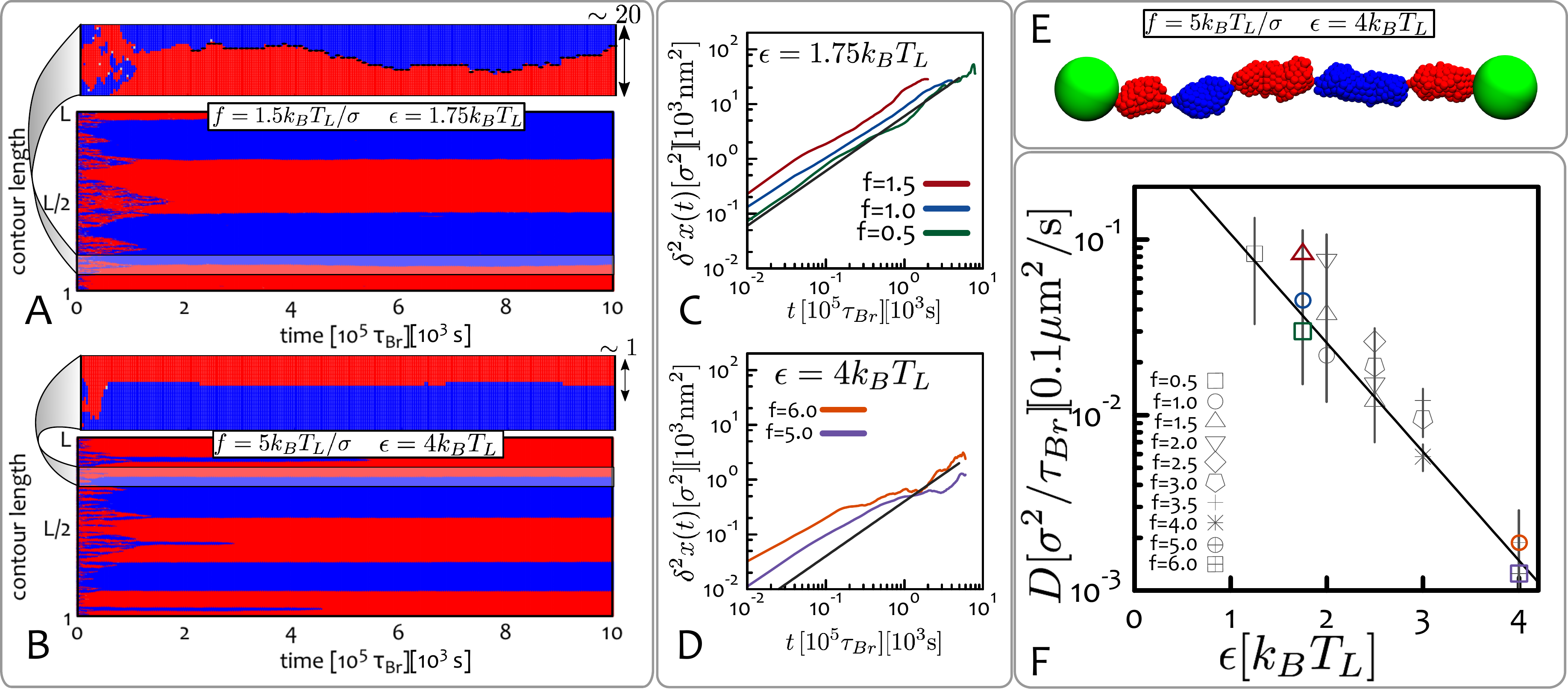}
 	\caption{\textbf{Epigenetic Domains and Boundary Diffusion.} (\textbf{A}-\textbf{B}) These panels show kymographs of systems displaying long-lived multi-domain states. Although these are not stable (i.e., free-energy minimising) states, they evolve on time-scales much longer than the simulations (or experiment) runtime. Zoomed in regions displayed at the top of each kymograph pinpoint the dynamics of boundaries between epigenetic domains. (\textbf{C}-\textbf{D}) These panels show typical mean-squared displacements of the boundaries $\delta^2 x(t)$ obtained from the kymographs and for a given $\epsilon$. \red{Curves are coloured corresponding to the indicated applied stretching force $f$}. (\textbf{E}) Typical snapshot of the system whose kymograph is reported in (\textbf{B}), i.e. $\epsilon=4 k_BT_L$ and $f=5 k_BT_L/\sigma$. (\textbf{F}) Plot of the diffusion coefficient $D$ extracted from $\delta^2 x(t)$. Data-points are grouped into symbols corresponding to equal stretching force $f$ (in units of $k_BT_L/\sigma$) and plotted as a function of $\epsilon$ in a log-linear plot to highlight the exponential decay. \red{Coloured symbols correspond to the combinations of $f$ and $\epsilon$ exemplified in (\textbf{C}) and (\textbf{D}).} Units of measure mapped to real length- and time-scales are shown. See also Supplementary movies. }
 	\label{fig:panel_kymo_2}
 \end{figure*}

In order to further quantify the behaviour of the system in parameter space, we systematically vary $f$ and $\epsilon$ and quantify the equilibrium states by measuring the values of the absolute ``epigenetic magnetisation''
\begin{equation}
m \equiv \dfrac{\left| n_{\rm red} - n_{\rm blue} \right|}{L}
\end{equation} 
and of the average chromatin extension via its radius of gyration 
\begin{equation}
R_g^2 = \dfrac{1}{L}\sum_{i=0}^L \left[ \bm{r}_{i} - \bm{r}_{\rm CM}\right]^2 \, ,
\end{equation}
where $\bm{r}_i$ and $\bm{r}_{\rm CM}$ are the positions of segment $i$ and of the centre of mass of the chain, respectively. In order to construct an equilibrium phase diagram of the system, we estimate the average $\langle R_g \rangle \equiv \langle R_g^2 \rangle^{1/2}$ and $\langle m \rangle$, first by time-averaging over a trajectory at steady state and then averaging the results across several trajectories. 
By measuring these two observables for different choices of $\epsilon$ and $f$ we report the phase diagram shown Fig.~\ref{fig:panel_kymo_pt}(C). One can readily notice that the typical equilibrium configurations can be grouped into two distinct phases: one is compact and epigenetically ordered (CO), i.e. with 
\begin{equation}
\langle R_g \rangle \sim L^{1/3}\sigma, \, \,  \langle m \rangle > 0 
\end{equation}
whereas the other is stretched and epigenetically disordered (SD), i.e. with 
\begin{equation}
\langle R_g \rangle \sim L \sigma, \, \, \langle m \rangle \simeq 0\, . 
\end{equation}
These two regimes are separated by a transition line and one can readily appreciate from the profiles of $\langle R_g \rangle$ and $\langle m \rangle$ (Figs.~\ref{fig:panel_kymo_pt}(D-E)) that this transition is abrupt.
In the limit of stretching forces $f \rightarrow 0$, we retrieve the first-order transition observed in Ref.~\cite{Michieletto2016prx} for un-stretched chromatin.
When larger stretching forces are applied, the system therefore displays a force-dependent critical line $\epsilon_c(f)$ (or $f_c(\epsilon)$) which retains the first-order-like features observed at $f=0$.

\begin{figure*}[t]
	\centering
	\includegraphics[width=0.8\textwidth]{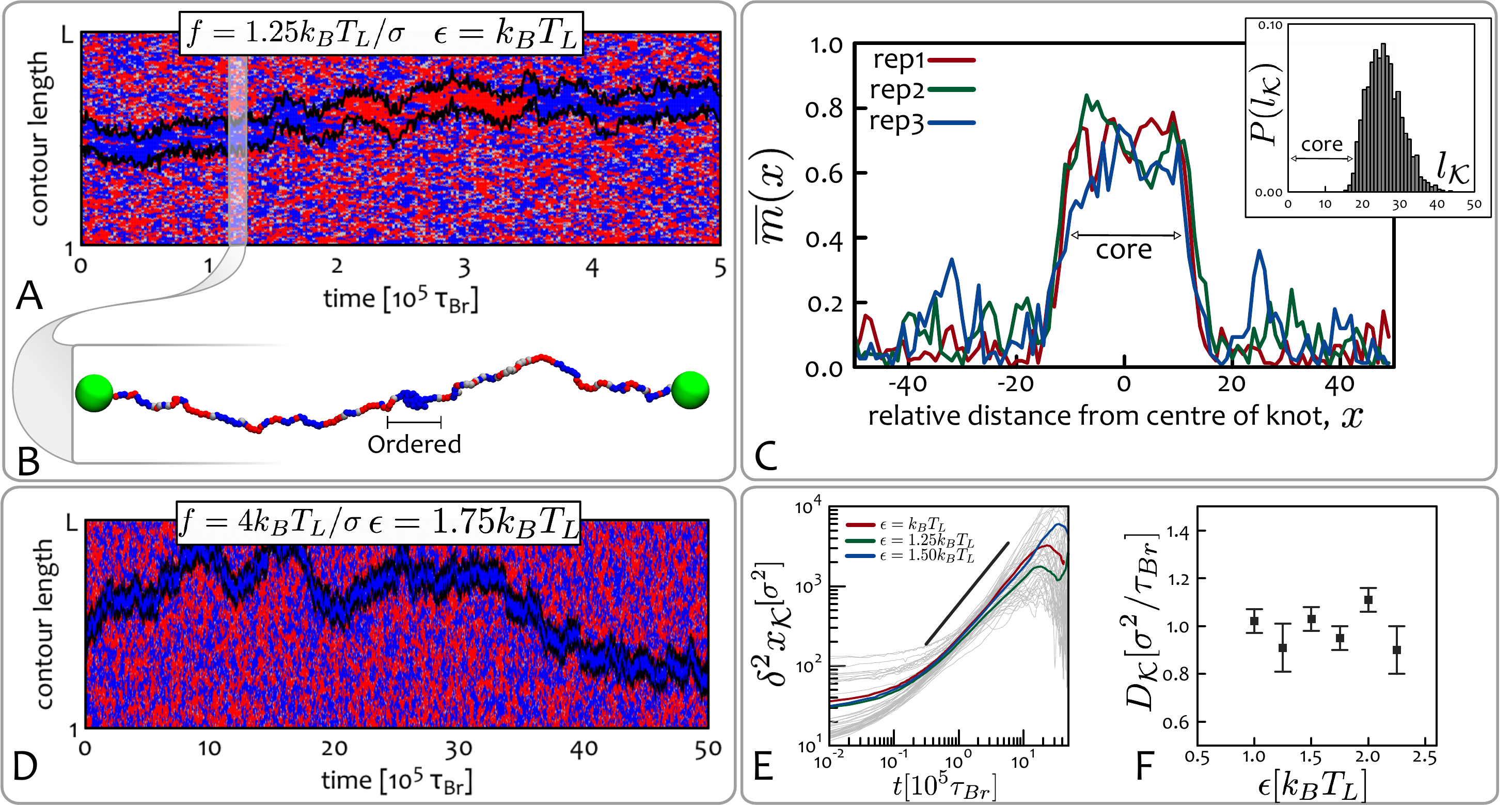}
	\caption{\textbf{Epigenetic knotted solitons.} (\textbf{A}) Kymograph corresponding to simulations initiated with a knotted chromatin fibre in equilibrium, with $f=1.25 k_BT_L/\sigma$. At time $t=0$ the system is flooded with reader and writer proteins having binding affinity $\epsilon=kBT_L$. The chromatin contains a figure-of-eight ($4_1$) knot~\cite{Adams1994}, whose escape from the fibre is avoided by the macroscopic beads at the terminal ends. Superimposed on the kymograph, we also show the boundaries of the knot computed from the 3D polymer configurations through the knot identification algorithm described in Ref.~\cite{Tubiana2011} (black lines). 
	Intriguingly, from this plot one can immediately realise that the knotted portion of the chain can be identified with the epigenetically ordered region, i.e. the chromatin segment where the beads are homogeneously coloured. (\textbf{B}) Snapshot of the chromatin fibre at the time-step highlighted in the kymograph. (\textbf{C}) Analysis of the knotted soliton for different system replicas. In the main panel we show the time-averaged magnetisation $\overline{m}(x)$ at position $x$ relative to the centre of the knot (see Materials and Methods); this displays a localised increase which we identify as an ``epigenetic soliton''. In the inset we show the probability of observing a knotted arc of length $l_\mathcal{K}$ through the knot identification algorithm of Ref.~\cite{Tubiana2011}. As one can notice, the length of the epigenetically ordered region and that of the knotted arc are in near-perfect agreement. This strongly suggests that the localised knotted segment leads to the local coherency in the epigenetic marks. \textbf{(D)} Increased affinity $\epsilon$ leads to slower switching times: here the epigenetic soliton never changes state during a 10 times longer simulation. \textbf{(E-F)} The measured mean squared displacement $\delta^2 x_\mathcal{K}(t)$ and diffusion coefficients $D_\mathcal{K}$ of the knotted solitons are shown to be insensitive to the interaction strength $\epsilon$. Here we consider a chromatin fibre with $L=200$ beads. See also Supplementary Movies.}
	\label{fig:panel_soliton}
\end{figure*}

Within the CO region, there is a parameter range where a multi-domain epigenetic structure emerges dynamically. This can be readily seen from the profile of the epigenetic magnetisation $\langle m \rangle$ in Fig.~\ref{fig:panel_kymo_pt}(E): near the transition line (from the compact ordered side) this quantity displays a sharp peak which then decreases for smaller values of the force. In the region labelled as ``multi'' in Fig.~\ref{fig:panel_kymo_pt}(E), the magnetisation is lower than unity because multiple macroscopic ordered regions populated by different epigenetic marks compete with one another, thereby lowering the overall magnetisation. These domains must be metastable (though long-lived) as in steady state a single domain is preferred since it removes domain walls, which have a free energy cost. The existence of multi-domain patterns arises because, at stretching forces far from critical $f_c(\epsilon)$, there may be multiple nucleation points along the fibre for the spreading of epigenetic marks. These nucleation points are generated by a local increase in chromatin density, for instance through the transient formation of loops and 3D interactions, which then trigger local spreading of different epigenetic marks (see Fig.~\ref{fig:panel_kymo_2}). As the stretching force gets closer to the critical value $f_c(\epsilon)$, the nucleation probability declines, hence only one epigenetic mark is able to take over the whole chromatin strand, thereby enhancing its overall epigenetic magnetisation.   

It is useful to compare the epigenetic transition between the CO and the SD state in Fig.~\ref{fig:panel_kymo_pt} with the equilibrium transition between a compact and a stretched state which can be observed in a stretched homopolymer with self-attractive interactions (i.e., in a poor solvent). While the transition is first order in the homopolymer case as well~\cite{homopolymer,homopolymer2}, here the epigenetic degrees of freedom increase the entropy of the disordered phase, thereby the value of the critical force, for a given $\epsilon$, is lower. Another key difference is that multi-pearl structures are only observed transiently close to the transition for the homopolymer case~\cite{homopolymer}, whereas in the epigenetic case multi-domain states arise far from the transition, and they are long-lived. Both these differences should be experimentally detectable, as the homopolymer case may be recreated by reconstituting chromatin fibres with histone octamers with controlled biochemical tags or in the presence of linker histones (H1), while avoiding the action of writers.

\subsection{Long-Lived Epigenetic Domains and Boundary Diffusion}
The competition of epigenetic marks over a chromatin segment is a phenomenon relevant to many biological systems. For example, genes that are positioned near telomeres may switch between transcriptionally active and silenced states from one generation to the next~\cite{Gottschling1990}. This phenomenon, generally referred to as ``telomeric position effect'', is particularly relevant in yeast~\cite{Gottschling1990} but it has also been observed in human cells~\cite{Baur2001}. It occurs because heterochromatin marks largely populating the telomeric gene-poor regions of the genome invade gene-rich stretches of chromatin, effectively switching off gene transcription. 

In our proposed single molecule set-up, the emergence of multiple competing epigenetic domains in reconstituted chromatin under (small) tension allows a quantitative study of stability and diffusive dynamics of boundaries between neighbouring epigenetic domains, processes which may be relevant to understand the mechanisms underlying competition and spreading of epigenetic marks {\it in vivo}.

\begin{figure*}[t]
	\centering
	\includegraphics[width=0.9\textwidth]{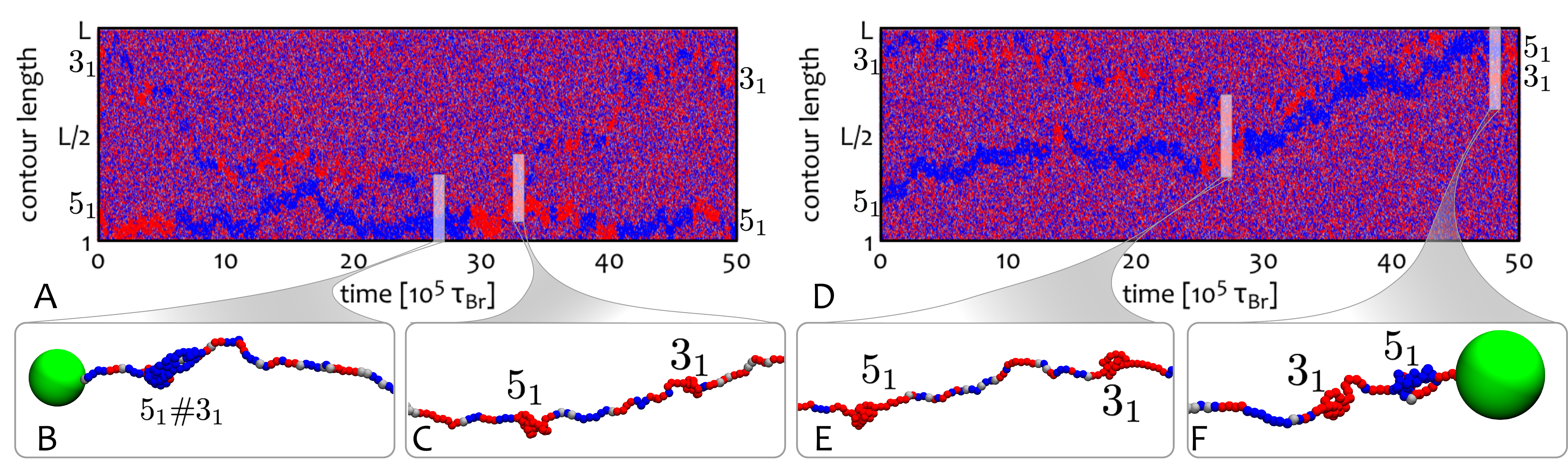}
	\caption{\textbf{Interacting Solitons.} The kymographs display reflecting \textbf{(A)} or cross-through~\cite{Trefz2014} \textbf{(D)} interactions between epigenetic knotted solitons. Panels \textbf{(B-C)} and \textbf{(E-F)} show typical situations where knots can form a unique composite knot or form independent domains. Here the chromatin strand has length $L=400$ beads.}
	\label{fig:panel_2knots}
\end{figure*}

In practice, an epigenetic domain in the stretched chromatin fibre is defined as a macroscopic region (e.g., consisting of $w\ge 50$ beads) over which more than $\red{\theta} = 90\%$ of the beads are homogeneously coloured. A boundary between domains is found if the beads on either side of the boundary have opposite colours (i.e., red and blue) and the signed magnetisation difference $\Delta \tilde{m}$ of the left and right domain is such that $\left| \Delta \tilde{m} \right| >2 \red{\theta}$ (i.e. the domains have opposite signed magnetisation). 

By tracking the position of the domain boundaries over time, one can measure their mean square displacement
\begin{equation}
	\delta^2 x(t) = \dfrac{1}{T-t} \int_0^{T-t} \left[ x(t_0+t) - x(t_0) \right]^2  dt_0 
	\label{eq:msd}
\end{equation} 
where $x(t)$ is the position of a given boundary at time $t$ \red{and $T$ the total measurement time}.  Plots of $\delta^2 x(t)$ for different replicas and two choices of interaction energies $\epsilon$ are reported in Fig.~\ref{fig:panel_kymo_2}(C-D). Data points are grouped according to the applied stretching force $f$, showing that the dependence on $f$ is much weaker compared to that on $\epsilon$. We extract the diffusion coefficient $D$ as the long time behaviour of $\delta^2 x(t)$, i.e.  
\begin{equation}\label{eq:D}
	D = \lim_{t \rightarrow \infty} \dfrac{\delta^2 x(t)}{2 t} \, ,
\end{equation}  
which is reported in Fig.~\ref{fig:panel_kymo_2}(F) and shows an exponential decay of $D$ as a function of $\epsilon$. 

In our stretched chromatin assay, the diffusive dynamics of epigenetic boundaries is therefore mainly controlled by the strength of the attractive interactions between beads bearing the same epigenetic mark. {\it In vivo}, another possible important factors affecting diffusivity of epigenetic boundaries may be the presence of insulators and architectural proteins such as CTCF and cohesins~\cite{Alberts2014,Stigler2016,Kanke2016}. These features may be included in future studies focused on understanding the actions of these architectural elements.



\subsection{Epigenetic Knotted Solitons}
The abrupt, first order, phase transition between the compact-ordered and the stretched-disordered state shown in Figure~\ref{fig:panel_kymo_pt} is here due to the coupling between {\it global} 3D structure and 1D epigenetic dynamics~\cite{Michieletto2016prx}, which gives rise to a positive feedback loop where compaction aids spreading, which leads to further crumpling at small enough $f$, or sufficiently large $\epsilon$.

It is interesting to ask whether one may design an experimentally realisable situation whereby this positive feedback is only realised {\it locally}, rather than globally as done in Ref.~\cite{Michieletto2016prx}. To do so, we consider a knotted chromatin fibre: the idea behind this construct is that a polymeric knot tightly localises upon stretching~\cite{Farago2002,Caraglio2015}, and that the knotted region is more compact, thereby providing a natural nucleation point for epigenetic ordering. In the single molecule experiment set-up we consider (Fig.~\ref{fig:panel_kymo_pt}), a physical knot may be tied within chromatin by using, for instance, micro-manipulating techniques used to generate knots on DNA~\cite{arai1999tying,Bao2003,Liu2016}. Alternatively, a knotted chromatin fibre may be self-assembled by first knotting the naked DNA strand, and later allowing nucleosome formation by adding histone octamers to the solution~\cite{Brackley2015nar}. 

In Figure~\ref{fig:panel_soliton} (A-B) we show the kymograph for a chromatin strand along which a figure-of-eight ($4_1$) knot is tied. As one can see, even though we choose values of $\epsilon$ and $f$ so that the system is in the stretched-disordered region of the phase diagram (see Fig.~\ref{fig:panel_kymo_pt}), the kymograph clearly displays a localised epigenetically ordered region. To confirm whether this region corresponds to the knotted arc, tightened upon stretching and self-attraction, we pinpoint and monitor the time evolution of the knotted region by resorting to a top-down search of the smallest portion of chromatin that yields, upon suitable closure, a ring with the same topology of the whole chain~\cite{Tubiana2011}. Remarkably, the knotted portion found with this well-established algorithm perfectly matches the epigenetic domain in the kymograph (black lines in Fig.~\ref{fig:panel_soliton} identify the boundary of the knotted arc, see also Methods). 

We call this remarkable configuration an ``epigenetic knotted soliton'', because, as a soliton, it travels along the fibre by keeping near-constant shape, hence displaying particle-like behaviour. Similarly to solitons recently observed in meta-materials~\cite{Chen2014} and minimal surfaces~\cite{Machon2016}, it is the non-trivial topology of the localised knot which keeps the structure together. The epigenetic state of the knotted soliton can ``switch'' from red to blue, or vice versa -- switching events occur, in Figure~\ref{fig:panel_soliton}, at about $2$ $10^5 \tau_{Br}$ and $3.5$ $10^5 \tau_{Br}$). Switching between the two stable states is observed because the soliton has a finite size, hence the effective free energy barrier separating the red and blue states can be occasionally overtaken by fluctuations. 


The physical properties of the soliton are tunable: for instance, by increasing the stretching force the knot tightens and becomes more compact, so that the soliton shrinks in size but becomes more stable due to the increase of 3D interactions which locally ``protect'' the epigenetically ordered state. By increasing values of $\epsilon$, such that the system remains overall disordered ($\epsilon<\epsilon_c(f)$) the switching rate decreases (i.e., the epigenetic ordering is more robust, see Fig.~\ref{fig:panel_soliton}(D)) -- the diffusion coefficient, however, does not depend on $\epsilon$ appreciably (unlike for the case of epigenetic domains, see Fig.~\ref{fig:panel_kymo_2}(E-F)).
It would be of interest to quantify how the soliton diffusion coefficient depends on knot type, and whether, as in knots in swollen polymers, twist knots and more complex knots are less mobile~\cite{Adams1994,Bao2003,diStefano2014,matthews2010}.  

For values of $\epsilon > \epsilon_c(f)$, we observe the same SD-CO phase transition reported in Figure~\ref{fig:panel_kymo_pt}, where the localised knot now acts as a nucleation point for the spreading of its epigenetic mark. 

Finally, we investigate how knotted solitons interact on a chromatin strand. We consider a system where two knots, a pentafoil $5_1$ and a trefoil $3_1$, are initially tied along the chromatin. As the system evolves each knot forms its independent epigenetic soliton. As shown by the kymographs in Figure~\ref{fig:panel_2knots}(A-D), both solitons diffuse along the chain~\cite{Bao2003}; they also interact and may merge to form a single domain, and later on split again (Fig.~\ref{fig:panel_2knots}(A)), or even cross each other~\cite{Trefz2014} (Fig.~\ref{fig:panel_2knots}(D)). Following every collision or interaction, each soliton retains its overall structure and re-establishes its locally coherent epigenetic mark.

\begin{figure}[t]
	\centering
	\includegraphics[width=0.45\textwidth]{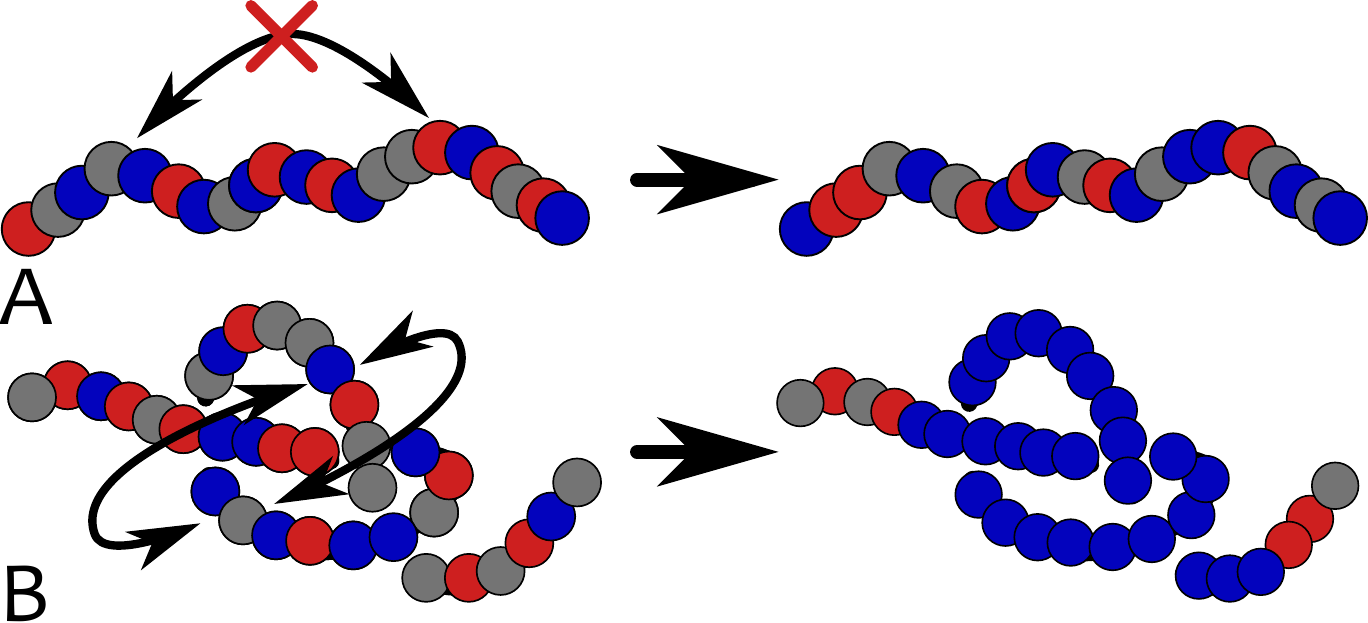}
	\caption{\textbf{Epigenetics and 3D topology.} This figure intuitively shows the reason behind the formation of a stable epigenetic domain within the knotted chromatin segment: the local 3D topology of a knot ``protects'' the underlying epigenetic information by enhancing local 3D interactions (\textbf{B}). On the contrary, an unknotted stretched segment lacks long-range interactions required to sustain epigenetic coherency (\textbf{A}). }
	\label{fig:panel_final}
\end{figure}

\section{Discussion and conclusions}
In several areas of chromosome biology, researchers assume the existence of a tight coupling between 3D chromatin structure and epigenetic dynamics, as this coupling provides an appealing mechanism for the {\it de novo} establishement and maintainance of epigenetic patterns~\cite{Dodd2007,Michieletto2016prx,Canzio2013,Angel2011,Ciabrelli2017}. However, as both epigenetic spreading and chromatin dynamics may occur on comparable and relatively fast timescales {\it in vivo} (minutes to hours~\cite{Zentner2013,Barth2010}), it is extremely difficult to design an experiment to demonstrate this coupling dynamically inside the cell. Here, instead, we proposed, and investigated {\it in silico}, a set-up for an experiment to test and quantify the coupling directly {\it in vitro}. The set-up involves a reconstituted chromatin fibre~\cite{Kruithof2009,Cui2000}, which is stretched (e.g., by laser tweezers) and interacts with an ensemble of reader and writer proteins -- such as heterochromatin HP1 and SUV39h1 complexes~\cite{Narlikar1,Canzio2013} or polycomb~\cite{Angel2011,Ciabrelli2017} enzymes able to bind to and deposit their respective marks.

We showed that by varying, for instance, the stretching force, one can trigger a phase transition between a compact-ordered phase, where a single epigenetic mark invades the whole chromatin fibre, and a stretched-disordered phase, where no single mark can take over the system. It is important that this transition may be observed by varying only the stretching force, as this is simpler to fine tune continuously {\it in vitro} with respect to temperature (which can inactivate readers or writers altogether) and effective self-attraction between chromatin segments (which may be achieved through the use of mutant reader proteins). 

By using single-molecule imaging~\cite{Yardimci2012,Kanke2016,Stigler2016} or super-resolution techniques~\cite{Huang2008} 
one may envisage to employ antibodies recognising H3K9me3 and H3K27me3 epigenetic states to separately stain the marks and thereby discriminate between an epigenetically ordered and an epigenetically disordered state. In principle, one may even be able to observe the transition from one to the other by increasing the stretching force in the single molecule assay. Another possibility to assess the degree of epigenetic order is to use single-cell chromatin immunoprecipitation~\cite{Rotem2015} (ChIP) on the reconstituted chromatin, to quantify histone modifications along the fibres. Note that, for this approach to be viable, one would have to use genomic DNA to reconstitute chromatin (rather than repeating, such as 601, sequences), so that locations along the DNA can be mapped uniquely. Therefore, whilst an experimental investigation of the compact-ordered to stretched-disordered transition is certainly challenging, and will require state-of-the-art experimental techniques, it is in principle feasible, and would constitute the first direct measurement of the so far elusive coupling between chromatin structure and epigenetic dynamics.
\red{A more experimentally accessible insight (although less informative regarding the epigenetic state of the system) may also be achieved via optical-tweezers by measuring force-extension curves~\cite{Kruithof2009,Cui2000,VanDenBroek2010}. These can discriminate between a (partially) collapsed and a stretched coil~\cite{VanDenBroek2010}, while cannot determine whether the chromatin is epigenetically ordered or disordered. On the other hand, the typical force-extension curves obtained with this set-up may display sensitivity on the concentration or efficiency of epigenetic readers and writers thereby offering indirect quantification of the recolouring process}.

By tying a knot along the reconstituted chromatin fibre we observed that the system can harbour a novel structure, an ``epigenetically knotted soliton'' (see Figs.~\ref{fig:panel_soliton}-~\ref{fig:panel_final}). This is a tight knot which is locally ordered epigenetically and diffuses freely, within an epigenetically disordered background. The size of the soliton is tunable by varying the stretching force applied to the chromatin fibre, and different solitons interact in a variety of ways: they may bounce off one another or cross through each other when they collide.
We also expect similar topological solitons to be universally found in knotted magnetic polymer, a new kind of topological soft matter which has not yet been realised experimentally~\cite{Garel1999a}. 

The findings we have reported here may also be employed to \emph{detect} knots in chromatin strands. Starting from an ordered and crumpled state (such as heterochromatin~\cite{Alberts2014}), one may imagine to apply and continuously increase an external stretching force $f$ and monitor the evolution of the system; whereas ordered unknotted regions will undergo an abrupt phase transition and become epigenetically disordered (see phase diagram in Fig.~\ref{fig:panel_kymo_pt}), knotted chromatin will instead preserve a localised epigenetically ordered region that can be identified with the knotted segment (Fig.~\ref{fig:panel_soliton}). 

Besides all this, the epigenetic solitons provide another observable consequence of the dynamic coupling between 3D chromatin structure and epigenetics which can be tested in future single-molecule experiments.

\section{Methods}\label{sec:ModMethods}
The datasets generated during and/or analysed during the current study are available from the corresponding author on reasonable request.

\subsection{Chromatin Model}
Chromatin is modelled as a bead-spring polymer chain~\cite{Kremer1990} where each bead has nominal size $\sigma$. \red{Attractive and repulsive interactions are controlled by the potential described in the main text (Eq.~\eqref{eq:pot}) and plotted in Fig.~S1. This is a truncated-and-shifted Lennard-Jones potential which is broadly used in Molecular Dynamics~\cite{FrenkelMD} to model short-ranged interactions. We include a colour-dependence in order to model like-colour attraction and different-colour repulsion as described in eq.~\eqref{eq:pot} and shown in Fig.~S1}.
Harmonic springs between beads $a$ and $b$ are imposed as $U^{ab}_{H}(x)= (\delta_{b,a+1} + \delta_{b,a-1})k_BT_L\kappa(x-x_0)/2 $ with $x_0=1.1\sigma$ and $\kappa=100 k_BT_L$ to ensure the connectivity of the backbone. The chain stiffness is regulated by a Kratky-Porod potential between triplets of beads forming an angle $\theta=\left(\bm{t}_a \cdot \bm{t}_b\right)/\left( \left| \bm{t}_a \right| \left| \bm{t}_b\right|\right)$ where $\bm{t}_a$ is the vector joining beads $a$ and $a+1$ as $U^{ab}_{KP}(\theta) = (\delta_{b,a+1} + \delta_{b,a-1}) k_BT_L l_p\left(1+\cos{\theta}\right)/\sigma$ with $l_p=3\sigma$.
The total potential $U^a(x)$ experienced by each bead is
given by the sum over all the possible interacting pairs and triplets, i.e.
\begin{equation}
Ua(x) = \sum_b \left[ U^{ab}_{LJ}(x)+U^{ab}_{H}(x)+U^{ab}_{KP}(x) \right] .
\end{equation} 
The dynamics of each bead therefore obeys the Langevin equation
\begin{equation}
	m\dfrac{d^2 \bm{r}_a}{dt^2} = -\gamma \dfrac{d \bm{r}_a}{dt} - \nabla U^a(x) + \bm{\xi}^a(t)  \label{eq:Lang}
\end{equation}
where $\gamma=1$ (in dimensionless LJ units) is a friction coefficient and $\bm{\xi}^a(t)$ is stochastic delta-correlated noise which obeys the fluctuation dissipation relationship
$\langle \xi_{\alpha,a}(t)\xi_{\beta,b}(t^\prime) \rangle= 2\gamma k_BT_L\delta(t -t^\prime)\delta_{\alpha \beta}\delta_{ab}$ where the Latin indexes run over particles and Greek indexes over Cartesian components. \red{The mass of the beads are taken to be unity in the dimensionless Lennard-Jones units~\cite{FrenkelMD}}. \red{The beads employed at the ends of the chromatin are five times larger than the beads forming the polymer and have the same mass. In order to simulate the stretching of the chromatin, we directly apply a force $f$ on these beads.}  \red{Finally, Eq.~\eqref{eq:Lang} is integrated with a velocity-Verlet scheme within the LAMMPS~\cite{Plimpton1995} engine.}

Simulations units can be mapped to real ones by considering that the polymer beads can coarse-grain groups of nucleosomes. The nominal size of the beads can is here considered to be 30 nm, as capturing the thickness of reconstituted chromatin fibre~\red{\cite{Thoma1979,Widom1985,Robinson2006,Kruithof2009}}. The typical timescales over which each bead diffuses its own size is therefore 
\begin{equation}
\tau_{Br} = \dfrac{\sigma^2}{D_{\rm self}} = \dfrac{3 \pi \eta \sigma^3}{k_BT_L}\simeq 0.01 \text{s} \, ,
\label{eq:tbr}
\end{equation} 
where we used the Einstein relation $D_{\rm self}=k_BT_L/3\pi \eta \sigma$ and we considered $\eta=150 cP$ as the effective viscosity. 

The recolouring process occurs at inverse rate $\tau_R=10^3 \tau_{Br}$. In each attempt, a bead is selected at random and a colour change is proposed; the move is then accepted if satisfying the Metropolis criterion as described in the text. In other words, on average, all beads will have been attempted to change their colour every $\tau_R=10^3 \tau_{Br}$ steps.

For simplicity, we model reader and writer proteins implicitly. This is effectively done by setting attractive pair-wise interactions between beads bearing the same colour and by assuming a uniform selection probability during the recolouring attempts. In other words, the physical proximity of reader and writers is not required for bridging segments and changing the beads colour. This model is therefore a good approximation for the case in which reader and writer proteins saturate the solution. On the other hand, lowering the concentration of reader and writer proteins would effectively slow down the kinetics towards the equilibrium state: this situation is more complicated and we defer it to future studies. 

\subsection{Knotted Soliton}
 To test the stability of the knotted soliton we consider the time-averaged degree of epigenetic coherence  $\overline{m}(x)$ at position $x$ relative to the centre of knotted arc, $c_\mathcal{K}$. This can be defined as
  \begin{equation}
  \overline{m}(x) = \dfrac{1}{\Delta \tau}\int_{\tau_1}^{\tau_1+\Delta \tau} m(l,t)\delta(l - (c_\mathcal{K} + x)) dt,
  \label{eq:tave_magn}
  \end{equation}
  where $\Delta \tau$ is the typical time between epigenetic ``switching'', where the magnetisation of the whole knot changes sign (such as the one occurring in Fig.~\ref{fig:panel_soliton} at about $2$ $10^5 \tau_{Br}$ and $3.5$ $10^5 \tau_{Br}$) and $m(l,t)$ is the magnetisation at position $l$ and time $t$. In Fig.~\ref{fig:panel_soliton}(C) we report $\overline{m}(x)$ for three system replicas and compare it with the distribution of knot sizes obtained from the 3D polymer configurations (inset). As one can notice, the length of the ordered domain agrees with the length of the knotted arc. Further, the shape of the magnetisation profile $\overline{m}(x)$ displays a broad peak, which is characteristic of soliton-type solutions.

\bibliographystyle{nar}

\subsection{Acknowledgements}
We acknowledge ERC for funding (Consolidator Grant THREEDCELLPHYSICS, Ref. 648050). The authors would like to thank Nick Gilbert and Jim Allan for insightful discussions and useful remarks on the manuscript.

\subsection{Author contributions statement }
All authors conceived the research. D.Mi. performed the simulations and analysed the data. All authors wrote and reviewed the manuscript. 

\subsection{Additional Information}
The authors declare no conflict of interest.
%
%
%
%
%
%

%
%

\end{document}